\documentclass[aps,prl,notitlepage]{revtex4-1}
\usepackage{amssymb}
\usepackage{amsmath}
\usepackage{dsfont}
\usepackage{color}
\usepackage{graphicx}

\begin{document}
\title{Matter-wave dark solitons and their excitation spectra \\
in spin-orbit coupled Bose-Einstein condensates}

\author{V.\ Achilleos}
\email{vassosnbi@gmail.com}
\affiliation{Department of Physics, University of Athens, Panepistimiopolis, Zografos, Athens 157 84, Greece}
\author{J. Stockhofe}
\email{jstockho@physnet.uni-hamburg.de}
\affiliation{Zentrum f\"ur Optische Quantentechnologien, Universit\"at Hamburg, Luruper Chaussee 149, 22761 Hamburg, Germany}
\author{P.\ G.\ Kevrekidis}
\affiliation{Department of of Mathematics and Statistics, University of Massachusetts, Amherst, MA 01003-9305, USA}
\author{D.\ J.\ Frantzeskakis}
\affiliation{Department of Physics, University of Athens, Panepistimiopolis, Zografos, Athens 157 84, Greece}
\author{P. Schmelcher}
\affiliation{Zentrum f\"ur Optische Quantentechnologien, Universit\"at Hamburg, Luruper Chaussee 149, 22761 Hamburg, Germany}
\affiliation{The Hamburg Centre for Ultrafast Imaging, Luruper Chaussee 149, 22761 Hamburg, Germany}

\begin{abstract}
We present three  
types of dark solitons in quasi-one-dimensional
spin-orbit coupled repulsive Bose-Einstein condensates.
Among these families,
two are always stable, 
while the third one is only stable sufficiently close to the linear regime. 
The solitons' excitation spectra reveal the potential 
existence of a {\it second} anomalous mode. While the  
first such mode describes the soliton oscillatory motion
in a parabolic trap,  
the second, when present, reflects the double well structure of the underlying
single-particle spectrum. 
This novel mode results in moving density
stripes in the vicinity of the soliton 
core, 
or in an out-of-phase oscillation of the constituent 
components, with little effect on the 
nearly stationary striped total density of the composite soliton.
\end{abstract}

\maketitle
\section{Introduction}
The recent experimental
realization of spin-orbit coupling (SOC) in Bose-Einstein condensates (BECs)~\cite{spiel1,expcol,engels3}  and fermionic gases~\cite{fermi1,fermi2}
has stimulated considerable research interest. This is due to the fact that 
it paves the way towards a deeper understanding of exotic 
solid state systems such as topological insulators, 
making use of the highly controllable environment of cold atom physics~\cite{spiel3,hasan}.
Ultracold atomic gases promise to be an ideal system for the generation of versatile artificial gauge fields~\cite{dalib} 
and studying such Abelian or even non-Abelian gauge fields is relevant not only in mimicking solid state physics, but potentially even to fundamental theories such as quantum electro- or 
chromodynamics~\cite{qcd}.

In the above context, ground state properties of SOC-BECs, including phase separation 
and the existence of the stripe phase~\cite{ho,ho1}, as well as the collective dynamics 
and oscillations~\cite{coll}, have already been studied in some detail both in theory 
and experiments \cite{spiel1,expcol}. Furthermore, a relevant direction that has also 
attracted much attention is the study of topological excitations, such as skyrmions~\cite{skyrm}, 
vortices \cite{xuhan}, and Dirac monopoles \cite{dirac}; additionally, dark solitons 
in toroidal geometry \cite{brand} and bright solitons in quasi one-dimensional (1D) 
attractive SOC-BECs \cite{vassos} have also been predicted. While these structures have been studied in 
single- and multi-component BECs \cite{review1,review}, 
SOC can significantly 
enrich their structural, stability and dynamical properties.

It is the purpose of this work to highlight the above by presenting and analyzing dark soliton 
states in SOC-BECs confined in a highly anisotropic (quasi-1D) parabolic trap. Employing 
a multiscale expansion method, we find three different dark soliton families, featuring 
either a constant or a spatially modulated background density; in the latter case, dark solitons 
occur as excited states on top of the stripe phase of SOC-BECs 
(we call these states ``stripe solitons''). 
We perform a Bogolyubov-de Gennes (BdG) linearization analysis of the above soliton families, showing that for our parameter values
constant background solitons are always stable, while stripe solitons are stable only close to the 
linear limit. The characteristic negative energy (anomalous) mode 
 associated with the oscillation frequency of solitons 
near the trap center is identified within the excitation spectrum 
and its eigenfrequency is determined analytically. Importantly, in certain parameter regions we also find 
a {\it second} anomalous mode, which does not exist in single- or multi-component 
BECs, and is only sustained due to the double well structure of the SOC 
single particle spectrum, featuring two minima. Exciting this mode, we observe 
intriguing dynamics: constant background solitons feature 
moving stripes in the vicinity of the soliton core, somewhat
reminiscent of the Kelvin mode modulation of vortex lines~\cite{kelvin}; 
in the case of stripe solitons, we observe an out-of-phase oscillation of the constituent 
solitons, which 
does not affect the stationary total density of the composite state. 

\section{The model and its analytical consideration}
We consider a SOC-BEC confined in a quasi-1D parabolic trap, with
longitudinal and transverse frequencies $\omega_x \ll \omega_{\perp}$.
In the framework of mean-field theory, this system can be 
described by the energy functional \cite{spiel1,ho,ho1}:
\begin{eqnarray}
\mathcal{E}\!=\! \mathbf{u}^{\dagger} \mathcal{H}_0 \mathbf{u} +
\frac{1}{2} \left( g_{11}|u|^4+g_{22}|\upsilon|^4
+2g_{12}|u|^2|\upsilon|^2 \right),
\label{hamfull}
\end{eqnarray}
where $\mathbf{u}\equiv (u, \upsilon)^T$, and the condensate wavefunctions
$u$ and $\upsilon$ are related (through suitable rotations~\cite{spiel1}) 
to the two pseudo-spin components of the BEC.
The single particle Hamiltonian $\mathcal{H}_0$ in Eq.~(\ref{hamfull}) reads:
$\mathcal{H}_0=\frac{1}{2m}(\hat{p}_x \mathds{1}+k_L\hat{\sigma}_z)^2+V_{\rm tr}(x) \mathds{1}+\Omega\hat{\sigma}_x+\delta\hat{\sigma}_z,$
where $\hat{p}_x=-i\hbar\partial_x$ is the momentum operator in the longitudinal direction,
$m$ is the atomic mass, $\hat{\sigma}_{x,z}$ are the Pauli matrices, $\mathds{1}$ is the $2 \times 2$ unit matrix,
$k_L$ is the wavenumber of the Raman laser which couples the states $u$ and $\upsilon$, 
$\delta$ is the detuning from Raman resonance, and $\Omega$ is the strength of the Raman coupling. 
Additionally, $V_{\rm tr}(x)$ is the external trapping potential, 
considered to assume the usual parabolic form: 
$V_{\rm tr}=(1/2)m\omega_x^2x^2$. Finally, the effectively 1D coupling constants $g_{ij}$, 
are given by $g_{ij}=2\hbar\omega_{\perp}\alpha_{ij}$, where $\alpha_{ij}$ 
are the s-wave scattering lengths (assumed to be positive). 
Measuring energy in units of $\hbar \omega_\perp$, length in units of the 
transverse harmonic oscillator length
$a_\perp=\sqrt{\hbar/(m \omega_\perp)}$, time in units of $\omega_\perp^{-1}$, and 
densities in units of the scattering length $\alpha_{11}$, we 
derive from Eq.~(\ref{hamfull}) the following dimensionless equations of motion:
\begin{eqnarray}
i\partial_t u &=& \left(-\frac{1}{2}\partial^2_x-ik_L\partial_x + V_{\rm tr}
+ |u |^2+\beta |\upsilon|^2  +\delta \right)u + \Omega\upsilon, 
\nonumber \\
i\partial_t\upsilon &=& \left( -\frac{1}{2}\partial^2_x+ik_L\partial_x +V_{\rm tr}
+\beta|u |^2 +\gamma |\upsilon |^2 -\delta \right)\upsilon + \Omega u,
\nonumber \\
\label{GP1}
\end{eqnarray}
where $\beta=\alpha_{12}/\alpha_{11}$,
$\gamma=\alpha_{22}/\alpha_{11}$, and we have used 
$k_L\rightarrow a_\perp k_L$, $\Omega \rightarrow \Omega/( \hbar\omega_\perp)$ 
and $\delta \rightarrow \delta/( \hbar\omega_\perp)$; 
the trapping potential in Eqs.~(\ref{GP1}) is now given by 
$V_{\rm tr}(x) = (1/2)\omega_{\rm tr}^2 x^2$, 
where $\omega_{\rm tr} \equiv \omega_x/\omega_\perp$. 
The stationary counterpart of Eqs.~(\ref{GP1}) is obtained by 
factorizing $\mathbf{u}(x,t) = \mathbf{u}(x) \exp(-i\mu t)$, where $\mu$ 
denotes the chemical potential. 

For the above parameters, below we use the following values, relevant to 
the $^{87}{\rm Rb} \;$ $5{\rm S}_{1/2}$, $F=1$ manifold \cite{spiel1}:  
Raman wavenumber $k_L=8$, 
normalized Raman coupling strength $\Omega \in [0, 100]$, and 
ratios of the scattering lengths $\alpha_{11} : \alpha_{12} : \alpha_{22} = 1: 0.995 : 0.995$. 
It is thus physically relevant to use 
$\gamma \approx 1$ while we will let $\beta$ be
a free parameter.

\begin{figure}[tbp]
\includegraphics[width=8.7cm]{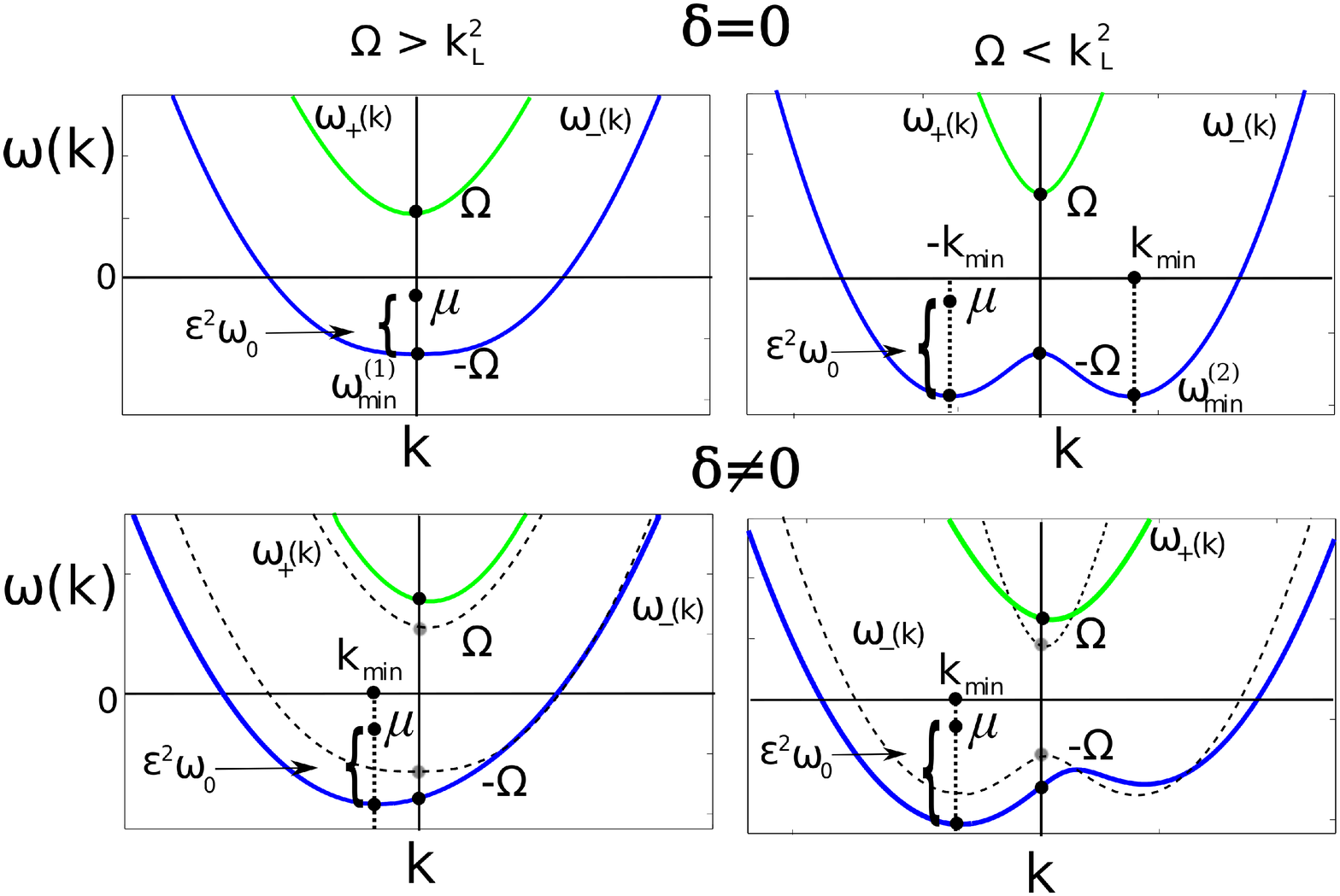}
\caption{(Color online) Sketches of the linear energy spectrum $\omega=\omega_{\pm}(k)$ 
for $\delta=0$ (top panels) and $\delta \ne 0$ (bottom panels), and also for 
for $\Omega> k_L^2$ (left panels) and for $\Omega< k_L^2$ (right panels). 
In the bottom panels, dashed lines depict (for comparison) the respective curves corresponding to $\delta=0$.
}
\label{fig1}
\end{figure}

We now use a multiscale perturbation method to derive approximate dark soliton solutions of Eqs.~(\ref{GP1}). 
This method \cite{kawa} utilizes proper scales and asymptotic expansions to reduce 
the original model into a simpler one; in our case, the latter will be a scalar nonlinear Schr\"odinger (NLS) equation, 
leading to dark soliton solutions.
To proceed, we 
introduce the ansatz: $\mathbf{u}= [U,V]^T \exp[i(kx-\mu t)]$, 
where $k$ is the momentum, 
$\mu=\omega+\epsilon^2\omega_0$ the chemical potential, 
$\omega$ is the energy in the linear limit, $\epsilon^2\omega_0$ 
is a small deviation about this energy 
($\epsilon \ll 1$), and $\omega_0/\omega=\mathcal{O}(1)$.  
We also assume that 
$\omega_{\rm tr}=\epsilon^2\tilde{\omega}_{\rm tr}$. 
Next, using 
the slow variables $T=\epsilon^2t$, $X=\epsilon x$, and 
the expansions 
$U=\sum_{n=1}^{+\infty} \epsilon^n U_n(X,T)$, 
$V=\sum_{n=1}^{+\infty} \epsilon^n V_n(X,T)$, 
we derive from Eqs.~(\ref{GP1}) the following equations at the orders 
$\mathcal{O}(\epsilon^1)$, $\mathcal{O}(\epsilon^2)$ and $\mathcal{O}(\epsilon^3)$, respectively:
\begin{eqnarray}
\!\!\! \mathbf{W}\mathbf{u}_1&=&0, \label{ord1} 
\\
\!\!\!\mathbf{W}\mathbf{u}_2&=&i\mathbf{W_0} \partial_X\mathbf{u}_{1}, \label{ord2} 
\\
\!\!\!\mathbf{W}\mathbf{u}_3&=&i\mathbf{W_0}\partial_X\mathbf{u}_{2} 
- \left(i\partial_{T}+\frac{1}{2}\partial_{X}^2 -\mathbf{A}+\omega_0\right)\mathbf{u}_{1},
\label{ord3}
\end{eqnarray}
where $\mathbf{u}_i=[U_i,V_i]^T$ ($i \in \{1, 2, 3\}$) are unknown vectors,
and matrices $\mathbf{W}$ and $\mathbf{A}$ are given by:
\begin{eqnarray}
\mathbf{W}&=&\left [ \begin{array}{c c}
\omega-k^2/2-kk_L-\delta & -\Omega \\
-\Omega & \omega-k^2/2+kk_L+\delta
\end{array}
\right], 
\nonumber \\
\mathbf{A}&=&\left [ \begin{array}{c c}
|U_1|^2 +\beta |V_1|^2 +\tilde{V}_{\rm tr}& 0 \\
0 & \beta|U_1|^2 + |V_1|^2+\tilde{V}_{\rm tr}
\end{array}
\right], 
\nonumber
\end{eqnarray}
where $\mathbf{W_0} = (\mathbf{W}- \omega \mathds{1})'$, 
primes denote differentiation with respect to $k$, and 
$\tilde{V}_{\rm tr}(X)=(1/2)\tilde{\omega}_{\rm tr}^2 X^2$. 
At $\mathcal{O}(\epsilon^1)$, the solvability condition 
${\rm det}\mathbf{W}=0$ yields the single-particle spectrum (dispersion
relation)
\begin{equation}
\omega = \omega_{\pm}(k) = \frac{1}{2} k^2 \pm\sqrt{(kk_L+\delta)^2+\Omega^2}, 
\label{dr}
\end{equation}
displayed in Fig~\ref{fig1}.
Let $\mathbf{L}=[1,Q]$ and $\mathbf{R}=[1,Q]^T$ 
be the left and right eigenvectors of $\mathbf{W}$ at eigenvalue $0$,  
where  
\begin{equation}
Q=Q(\omega,k) \equiv \frac{1}{\Omega}\left(\omega-\frac{1}{2}k^2 -kk_L-\delta\right). 
\label{Q}
\end{equation}
Then, the compatibility condition of Eq.~(\ref{ord1}) yields:
\begin{equation}
\mathbf{u}_1=\mathbf{R}\psi(X,T), 
\label{u1}
\end{equation}
where $\psi(X,T)$ is an unknown scalar field. 

Next, proceeding with Eq.~(\ref{ord2}), the compatibility condition 
$\mathbf{L}\mathbf{W_0}\mathbf{R}=0$ at the order $\mathcal{O}(\epsilon^2)$ 
enforces a vanishing group velocity
$v_g \equiv \partial \omega/\partial k$, {\it i.e.}:
\begin{equation}
v_g=k -k_L\frac{Q^2-1}{Q^2+1} = 0,
\label{vgr}
\end{equation}
which means that $\omega$ and $k$ should be 
evaluated at stationary points of the dispersion relation~(\ref{dr}).
In the following we focus on the energy {\it minima} $(\omega_{\rm min},k_{\rm min})$. 
Additionally, at the order $\mathcal{O}(\epsilon^2)$, we obtain the following solution for $\mathbf{u}_2$:
\begin{equation}
\mathbf{u}_2=-i\mathbf{R'}\partial_X\psi(X,T).
\label{u2}
\end{equation}
Finally, at $\mathcal{O}(\epsilon^3)$, the compatibility condition for Eq.~(\ref{ord3}), combined with 
Eq.~(\ref{u1}) and the above result for $\mathbf{u}_2$, yields the 
following NLS 
equation for the scalar field $\psi$:
\begin{eqnarray}
i\partial_T\psi+\frac{\Lambda}{2}\partial_{X}^2\psi- (\nu|\psi|^2-\omega_0)\psi=\tilde{V}_{\rm tr}(X) \psi,
\label{nls1}
\end{eqnarray}
where the coefficients $\Lambda$ and $\nu$ defined as:
\begin{eqnarray}
\Lambda= 1-\frac{2QQ'(k_L-k)}{1+Q^2}, \quad
\nu=\frac{Q^4+2\beta Q^2+1}{1+Q^2}, 
\label{nu}
\end{eqnarray}
are evaluated at the minima $(\omega,k)=(\omega_{\rm min}, k_{\rm min})$ 
of the dispersion relation (\ref{dr}).

In the homogeneous case 
($\tilde{\omega}_{\rm tr}=0$), for $\Lambda \nu >0$   
and 
boundary conditions $|\psi|^2 \rightarrow \omega_0/\nu$ 
as $|X| \rightarrow \infty$, the 
scalar NLS Eq.~(\ref{nls1}) 
possesses dark soliton solutions, $\psi_{\rm DS}$ (characterized by 
the free parameter $\omega_0$), of the form: 
\begin{eqnarray}
\psi_{\rm DS}=\sqrt{\omega_0/\nu}\left[\cos\theta{\rm tanh}(\eta)+i \sin\theta\right], 
\label{sol1} 
\end{eqnarray}
where $\eta=\sqrt{\omega_0/\Lambda}\cos\theta[X-X_0(T)]$, and 
$\theta$ is the 
``soliton phase angle'' 
($|\theta|<\pi/2$), 
$X_0(T)$ is the soliton center, 
while the soliton amplitude (depth) and soliton velocity are respectively given by 
$\sqrt{\omega_0/\nu}\cos\theta$ and 
$\dot{X}_0=\sqrt{\omega_0/\Lambda}\sin\theta$. 
Note that the limiting case $\theta=0$ corresponds to a stationary kink (``black'' soliton), while 
$\theta \ne 0$ give rise to travelling (``grey'') solitons; both types  
are identical to those found for the first time in the seminal work of Ref.~\cite{tsuzuki}.

\begin{figure}[tbp]
\includegraphics[width=8.5cm]{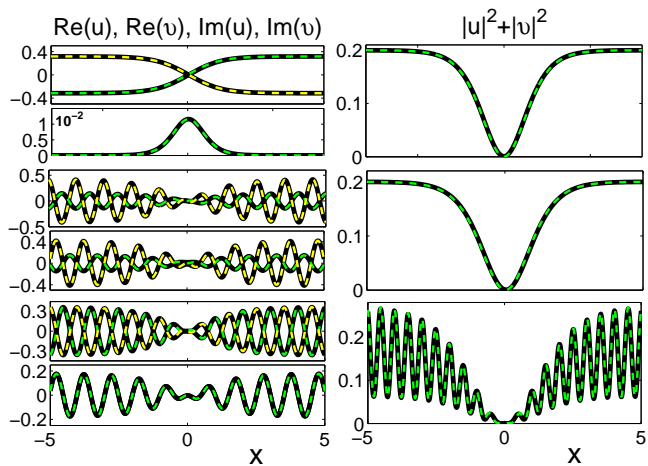}
\caption{(Color online) Each of the three triplets of panels shows 
the real (top left) and imaginary (bottom left) parts of $u$ and $\upsilon$, 
as well as total density profiles $|u|^2+|\upsilon|^2$ (right) of stationary kinks 
for $\delta=0$, $\beta=1$ and $\epsilon^2\omega_0=0.2$. 
Top triplet: 
a soliton in Regime~I for $\Omega/k_L^2=1.4$;
middle and bottom triplets: 
a $k_{\rm min}$-soliton and a stripe soliton in Regime~II for $\Omega/k_L^2=0.625$. 
Solid (black) lines depict numerical results, while dashed (green and yellow) lines 
depict analytical ones 
[cf.~Eqs.~(\ref{sol1a}), (\ref{sol3}) and (\ref{u2})]. }
\label{fig2}
\end{figure}%

Using the above expressions, we can now write down 
an approximate dark soliton solution of Eqs.~(\ref{GP1}), in terms of the original 
variables $x$ and $t$, as follows:
\begin{eqnarray}
\!\!\left(\!\!
\begin{array}{c}
u \\
\upsilon
\end{array}
\!\!\right) \approx \epsilon \psi_{\rm DS} \exp\left[ik_{\rm min}x -i(\omega_{\rm min}+\epsilon^2\omega_0)t \right]
\!\!\left(\!
\begin{array}{c}
\!\!1 \\
\!\!Q_{\rm min}
\end{array}\!\!
\!\right), 
\label{sol1a}
\end{eqnarray}
where $\psi_{\rm DS}$ is given in Eq.~(\ref{sol1}), but with 
argument $\eta \rightarrow \eta/\epsilon $, and 
$Q_{\rm min} \equiv Q(\omega_{\rm min}, k_{\rm min})$.

\section{Dark solitons in the homogeneous setting}
The coefficients of the NLS Eq.~(\ref{nls1}) and thus the soliton parameters, 
explicitly depend on $Q_{\rm min}$,  
which is calculated at the minimum of the dispersion relation (\ref{dr}). 
The latter possesses an upper branch $\omega_{+}$ and a lower branch $\omega_{-}$, as shown in Fig.~\ref{fig1}. 
We hereafter focus our analysis around the energy minima
of the lower branch $\omega_{-}$. 
We will study separately the regimes $\Omega/k_L^2>1$ (Regime~I) and $\Omega/k_L^2<1$ (Regime~II), 
for both $\delta=0$ and $\delta\ne 0$, which feature different characteristics regarding 
the energy minima, as shown in the left and right panels of Fig.~\ref{fig1}, respectively.

{\it Dark solitons in Regime~I.}
In this regime, and for $\delta=0$, the lower branch possesses a minimum 
$\omega_{\rm min}^{(1)}=-\Omega$ at zero momentum $k_{\rm min}=0$, while $Q_{\rm min}=-1$ 
and $Q_{\rm min}'=-k_L/\Omega$. The above values determine $\Lambda$ and $\nu$ in Eqs.~(\ref{nu}) and, thus, 
the form of the soliton in Eq.~(\ref{sol1a}). 
The existence of this soliton is numerically confirmed by solving the stationary 
version of Eqs.~(\ref{GP1}), 
where $\mu=-\Omega+\epsilon^2\omega_0$ [as per Eq.~(\ref{sol1})], using a fixed-point algorithm \cite{kelley}.
An example of a stationary kink ($\theta=0$), 
corresponding to 
$\Omega/k_L^2=1.4$ and $\epsilon^2\omega_0=0.2$, is shown
in the top triplet of panels of Fig.~\ref{fig2}. It is observed that 
the real parts of $u$ and $\upsilon$ (top left panel of the triplet) are opposite, 
in accordance with the form of the right eigenvector $\mathbf{R}=[1,-1]^T$; the imaginary parts
(bottom left panel of the triplet) are of order $\mathcal{O}(\epsilon^2)$, and are equal 
having a ${\rm sech}^2$ profile, in accordance to the analytical prediction for $\mathbf{u}_2$.
Finally, the spatial profile of the total density 
$|u|^2+|\upsilon|^2$ (top right panel) has the form of 
a scalar dark soliton's density \cite{review}. 

{\it Dark solitons in Regime~II.}
In this regime, and for $\delta=0$, we find solitons with energies 
near the energy minimum $\omega_{\rm min}^{(2)}$ 
at finite momenta,{\it i.e.},
$k_{\rm min}=\pm k_L (1-\Omega^2/k_L^4)^{1/2}$ (cf. top right panel of Fig.~\ref{fig1}). 
Hence,
$Q_{\rm min}^{(\pm)}=-k_{\rm min}\Omega^{-1}(k_L \pm k_{\rm min})$ and 
$Q_{\rm min}'^{(\pm)}=-\Omega^{-1}(k_L \pm k_{\rm min})$, with these values determining the
soliton parameters in Eq.~(\ref{nu}). 

It is thus clear that, in this regime, two different soliton solutions can be found, 
each corresponding to the locations $k = \pm k_{\rm min}$ 
of the energy minimum; these will be called hereafter $\pm k_{\rm min}$-solitons.
An example of a $k_{\rm min}$-soliton, for $\Omega/k_L^2=0.625$ and 
$\epsilon^2\omega_0=0.2$, is shown in the middle triplet of panels of Fig.~\ref{fig2}. 
As can be seen, the real and imaginary parts of $u$ and $\upsilon$ have different amplitudes 
in this case, due to the form of $\mathbf{R}=[1,Q_{\rm min}^{(\pm)}]^T$; furthermore, 
${\rm Re}(u, \upsilon)$ and ${\rm Im}(u, \upsilon)$ are now spatially 
oscillatory 
with a wavelength $2\pi/k_{\rm min}$. On the other hand, 
the total density profile features the usual ({\it i.e.}, unmodulated) density dip.

Importantly, still another branch of dark soliton solutions can 
be found in 
Regime~II as follows. First we note that the 
linear problem in this regime
admits solutions which are linear combinations of plane waves of momenta 
$k=\pm k_{\rm min}$. Then, employing continuation arguments, we construct approximate 
dark soliton solutions in the form of a linear combination of the above 
mentioned $\pm k_{\rm min}$-solitons. 
Solutions satisfying the symmetry $u=-\bar{\upsilon}$ 
(bar denotes complex conjugation) can be expressed as:
\begin{eqnarray}
\left( 
\begin{array}{c}
u \\
\upsilon
\end{array}
\right) \approx \epsilon C \psi_{\rm DS}
\!\! 
\left( 	
\begin{array}{c}
\!\!q_{+} \cos(k_{\rm min} x) 
+i q_{-} \sin(k_{\rm min} x) \\
\!\!-q_{+} \cos(k_{\rm min} x) 
+i q_{-}\sin(k_{\rm min} x)
\end{array}\!\!\!
\right)\!\! , 
\label{sol3}
\end{eqnarray}
where $q_{\pm} = \Omega^{-1}+Q_{\rm min}^{(\pm)}$ and $C$ is a free parameter. 
The existence of these solitons was also confirmed numerically, and a pertinent example 
is shown in the bottom triplet of panels of Fig.~\ref{fig2}, for $\Omega/k_L^2=0.625$, 
$\epsilon^2\omega_0=0.2$ and $C=0.8$. 
It is observed that the soliton background features a spatially modulated density, 
reminiscent of the stripe phase of the SO-coupled BECs \cite{ho,ho1}; for this reason, 
solitons of this branch will be called ``stripe solitons''. 

Both types of solitons in Regime~II were found to exist for any value of 
$\Omega/k_L^2$. 
The stripe- and $\pm k_{\rm min}$-solitons 
are excitations on top of their respective backgrounds, which are the ground states for small 
and large $\Omega$, respectively \cite{ho1}; thus, one expects 
stripe ($\pm k_{\rm min}$) solitons to be energetically preferable for small (large) 
$\Omega$ and the total density background to be modulated (uniform), accordingly. We have also confirmed (results not shown here) 
that solitons in Regime~II 
(or in Regime~I), could be created starting from the respective backgrounds, 
and using the experimentally relevant phase-imprinting method 
(i.e., utilizing an additional potential of the form of Eq.~(5) in Ref.~\cite{burger} 
to imprint a $\pi$-phase on the BEC). 

For all types of 
solitons 
in Regimes~I and~II, the 
results of Eqs.~(\ref{sol1a}) and (\ref{sol3}) are in an {\it excellent}
agreement with the numerically obtained solutions 
(see, respectively, dashed and solid lines 
in Fig.~\ref{fig2}). Furthermore, apart from the case of stationary (black) 
solitons, we have also studied travelling (grey) solitons 
with $\theta \ne 0$; this was done by direct numerical integration of Eqs.~(\ref{GP1}), 
by means of a Runge-Kutta method, using as initial conditions 
the analytical form of moving solitons, cf. Eq.~(\ref{sol1}). We have confirmed 
that near the linear limit ({\it i.e.}, for $\epsilon^2 \omega_0 \lesssim 0.1$) 
the solitons remain robust and evolve without distortion. 

{\it The case of nonzero detuning parameter.}  
Dark solitons can also be found for $\delta\ne 0$, upon 
calculating the energy minimum $(k_{\rm min}, \omega_{\rm min})$ of the lower branch 
of the energy spectrum~(\ref{dr}) (cf. bottom panels of Fig.~\ref{fig1})
and then using Eqs.~(\ref{nu}) to determine the soliton parameters for the solution~(\ref{sol1}). 
There are two important differences between the cases $\delta \ne 0$ and $\delta= 0$. 
First, since $k_{\rm min} \neq 0$ for every $\delta \neq 0$ (see Fig.~\ref{fig1}), 
solitons in Regime~I are shifted to finite $k$.
Second, since in Regime~II the degeneracy of the energy minima in the lower branch is always lifted 
for $\delta \ne 0$, stripe solitons cannot generically be constructed using the linear superposition argument 
as in Eq.~(\ref{sol3}).
We have numerically confirmed 
the existence of solitons  (resembling the middle triplet of panels of Fig.~\ref{fig2})
for $\delta \ne 0$ 
at the global minimum of the energy spectrum 
in both Regimes I and II.

\section{Stability and dynamics of solitons in the trap}
We now focus on 
the case $\tilde{\omega}_{\rm tr}\ne 0$, 
for which we have confirmed that all soliton families 
persist. Furthermore, for the stationary solutions, 
we have performed a linear stability analysis and identified regimes of (in)stability. Our analysis 
relies on the study of the BdG excitation spectrum around a stationary soliton solution  
${\bf u_{\rm sol}} \equiv (u_{\rm sol},v_{\rm sol})^T$ of Eqs.~(\ref{GP1}), 
with chemical potential $\mu$. 
The spectrum is obtained as follows. We introduce the ansatz
${\bf u} =\left\{ {\bf u_{\rm sol}} + \varepsilon [\exp(\lambda t) {\bf a}(x) 
+ \exp(\bar{\lambda} t) \bar{{\bf b}}(x)] \right\} \exp(-i \mu t)$,
where $\varepsilon$ is a formal small parameter, and 
$\{\lambda,({\bf a},{\bf b})\}$ define an eigenvalue-eigenvector pair. Then, 
substituting 
this ansatz into Eqs.~(\ref{GP1}) and linearizing, 
we arrive at $\mathcal{O}(\varepsilon)$ at 
an eigenvalue problem for eigenvectors $({\bf a},{\bf b})$ and eigenvalues $\lambda$. 
Note that as the latter may, in principle, be complex, {\it i.e.}, $\lambda=\lambda_r+i\lambda_i$,
instability corresponds to $\lambda_r >0$.

\begin{figure}[tbp]
\includegraphics[width=8.5cm]{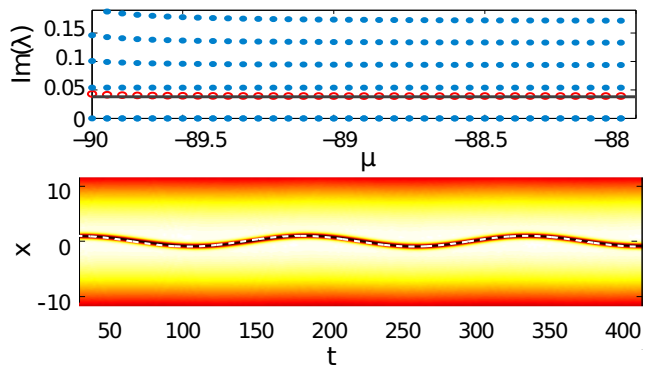}
\caption{(Color online) Top panel: 
the lowest imaginary eigenvalues of the 
linearization spectrum  
as functions of $\mu$, for a dark soliton in Regime~I, 
with $\delta=0$, $\Omega/k_L^2=1.4$, and $\tilde{\omega}_{\rm tr}=0.1$. 
Circles (in red) denote the anomalous mode eigenfrequency 
and the solid (black) line depicts the prediction of Eq.~(\ref{ode}). 
Bottom panel: contour plot showing the 
evolution of the density of a 
SOC-BEC carrying a 
dark soliton initially placed at $x_0=1$. Dashed (white) line
depicts the analytical result of Eq.~(\ref{ode}).
}
\label{fig4}
\end{figure}

First, we consider the case of solitons in Regime~I, for $\delta=0$, and study their  
stability starting from the linear limit, $\mu=-\Omega$, and entering into the nonlinear regime 
by increasing parameter $\omega_0$. 
This soliton branch is 
characterized by purely imaginary eigenvalues, thus it 
is dynamically stable. 
As shown  
in the top panel of Fig.~\ref{fig4} for $\Omega/k_L^2=1.4$ and $\omega_{\rm tr}=0.1$, there exists 
one anomalous (negative energy) mode, depicted by red circles, whose
frequency characterizes 
the small-amplitude oscillations of the dark soliton around the trap center. 
The latter can be obtained analytically in the framework of Eq.~(\ref{nls1}) as follows. 
As is well known, sufficiently deep (almost black) dark solitons oscillate in a parabolic 
trap of strength $\omega_{\rm tr}$ with a frequency 
$\omega_{\rm sol}=\omega_{\rm tr}/\sqrt{2}$ \cite{BA}; see also
the review of~\cite{review}. Hence, 
we can infer that solitons of Eq.~(\ref{nls1}) with $\theta\approx 0$ evolve in the trap 
so that their center $X_0(t)$ satisfies the equation of motion: 
\begin{eqnarray}
\frac{\text{d}^2 X_0}{\text{d}t^2}=-\frac{1}{2}\omega_{\rm sol}^2X_0, \quad
\omega_{\rm sol}=\sqrt{\frac{\Lambda}{2}}\omega_{\rm tr}.
\label{ode}
\end{eqnarray}
The above soliton oscillation frequency $\omega_{\rm sol}$ is   
depicted by the solid (black) line in the top panel of Fig.~\ref{fig4}. It can be observed 
that this analytical result almost coincides with the anomalous mode eigenfrequency. To 
further elaborate on this result, we have numerically integrated Eqs.~(\ref{GP1}) with 
an initial condition of a soliton with $\theta =0$, initially placed at $x_0=1$. As seen 
in the bottom panel of Fig.~\ref{fig4}, the soliton indeed performs 
harmonic oscillations
accurately described by Eq.~(\ref{ode}), cf. dashed (white) line in the figure. We note that, for $\delta\ne 0$,  
solitons have a spectrum qualitatively similar to the one in Fig.~\ref{fig4}, featuring one anomalous mode and no real ({\it i.e.}, unstable) eigenvalues.

\begin{figure}[tbp]
\includegraphics[width=8.5cm]{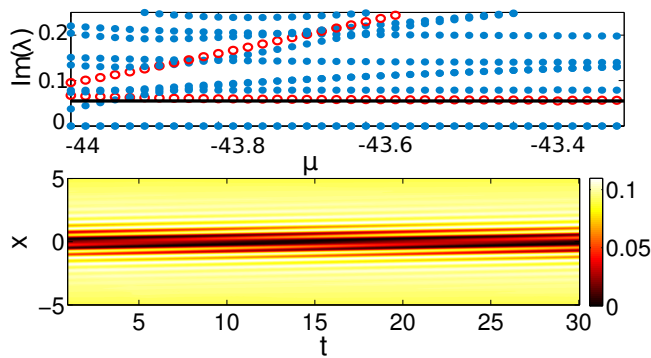}
\caption{(Color online) Top panel: same as the top panel of 
Fig.~\ref{fig4}, but for the $k_{\rm min}$-solitons in Regime~II with
$\Omega/k_L^2=0.625$. Bottom panel: contour plot showing the evolution of 
the density of the $u$-component, when perturbed by the eigenvectors of the second 
anomalous mode at $\mu = -43.5$. Other parameter values are the same as in Fig.~\ref{fig4}.}
\label{fig5}
\end{figure}

Next, we study the stability of solitons in Regime~II 
for $\delta=0$ and $\tilde{\omega}_{\rm tr}\ne 0$.
First we note that $k_{\rm min}$-solitons are stable, 
as no real eigenvalues appear in the spectrum -- see the example shown in the top panel 
of Fig.~\ref{fig5} for a soliton with $\Omega/k_L^2=0.625$. However, an important 
difference arises in this case, namely the emergence of a {\it second} anomalous 
mode in the linearization spectrum. Indeed, as shown in the figure [cf. circles (in red)], 
there exist two anomalous modes: a lower-lying one (with smaller eigenfrequencies), 
associated with the soliton oscillations as before, and an upper-lying anomalous mode (with larger eigenfrequencies). 
This second anomalous mode does not exist in the case of dark solitons in single- 
or multi-component BECs: its emergence is particular to dark solitons in SOC-BECs, 
which feature a double well structure in their energy spectrum in Regime~II. 
Indeed, in the linear limit, the system possesses two 
eigenstates that energetically lie below the $k_{\rm min}$-soliton, namely
one oscillator ground state configuration at each of the two minima of the dispersion relation.
Each of these gives rise to one anomalous mode in the soliton's linearization spectrum.
Superimposing the $k_{\rm min}$-soliton at momentum $k_{\rm min}>0$ with the oscillator
ground state at $k_{\rm min}>0$ leads to the familiar soliton oscillation, represented by 
the lower-lying anomalous mode. However, superimposing the 
$k_{\rm min}$-soliton at $k_{\rm min}>0$ with the oscillator ground state at $k_{\rm min}<0$, 
different dynamics occur:
adding counter-propagating 
waves of opposite $k$ gives rise to stripe signatures in the density.
We have numerically solved Eqs.~(\ref{GP1}), perturbed by the eigenvector 
$(\mathbf{a},\mathbf{b})$ of the second anomalous mode, to check this. 
The bottom panel of Fig.~\ref{fig5} shows the corresponding 
evolution of the density of 
$u$, the smaller component in our example where the stripe-forming effect is more evident.
Clearly, exciting the second 
anomalous mode indeed 
induces a periodic spatial modulation of the background, 
moving also through the soliton, with a wave number $\sim 2k_{\rm min}$.
This modulation effect inside the dark soliton core
is somewhat reminiscent of the way in which Kelvin modes modulate vortex 
lines~\cite{kelvin}.

\begin{figure}[tbp]
\includegraphics[width=8cm]{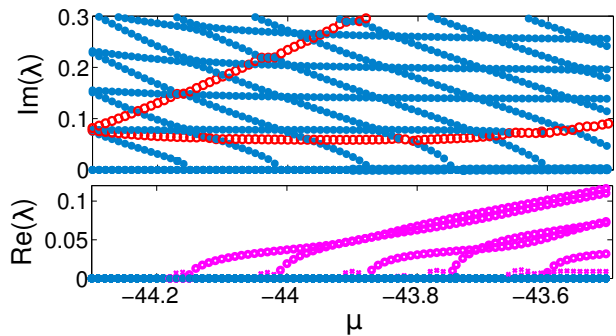}
\caption{(Color online) 
The imaginary (top) and the real (bottom) parts of the stability eigenvalues 
as functions of $\mu$, for a stripe 
soliton in Regime~II for $\Omega/k_L^2=0.625$. The 
soliton is stable, {\it i.e.}, 
${\rm Re}(\lambda)=0$, only near the 
linear regime ($\mu \lesssim -44.15$) and becomes unstable 
deeper in the nonlinear regime. 
}%
\label{fig6}
\end{figure}

\begin{figure}[tbp]
\includegraphics[width=8.5cm]{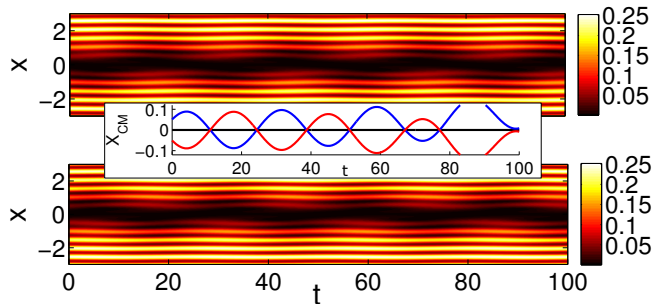}
\caption{(Color online) Same as in bottom panel of Fig.~\ref{fig5} but for both components of the 
stripe soliton in Regime~II with $\mu=-44$. The inset middle panel shows the 
time evolution of the center of mass $x_{\rm cm}$ of the $u$ component 
(initially at $x_{\rm cm}>0$ -- cf. blue line), of the $\upsilon$ component 
(initially at $x_{\rm cm}<0$ -- cf. red line), and of the total 
density (black line), 
at $x_{\rm cm}=0$. 
}
\label{fig7}
\end{figure}

Lastly, we consider stripe solitons in Regime~II [cf. Eq.~(\ref{sol3})]
for $\tilde{\omega}_{\rm tr}\ne 0$.
The BdG spectrum for such a solution 
with $\Omega/k_L^2=0.625$ is depicted in Fig.~\ref{fig6}. 
As expected,
the spectrum also features two anomalous modes, 
reflecting the existence of a doubly degenerate lower energy 
state. In the linear limit, the anomalous modes correspond to an in-phase and an out-of-phase superposition 
of the oscillator ground states, respectively. 
The former leads to in-phase oscillatory 
dynamics of the solitons in the two components, 
captured by the 
lower-lying anomalous mode.
On the other hand, the second anomalous mode corresponds to out-of-phase 
oscillations of the solitons. 
This is directly confirmed by propagation of an initially stationary stripe 
soliton, perturbed by the respective eigenvector, as shown in the contours of Fig.~\ref{fig7}.
The dark solitons in each component are shown to perform small-amplitude oscillations; 
the inset in Fig.~\ref{fig7} depicts the evolution of the center of mass 
for the individual components, clearly revealing that these are out-of-phase oscillations. 
Additionally, and contrary to $\pm k_{\rm min}$-solitons, 
Fig.~\ref{fig6} shows that at $\Omega/k_L^2=0.625$ stripe solitons have a small stability domain 
only close to the linear limit, and become unstable deeper in the nonlinear regime; 
the instability, though, is found to be induced by the background
({\it i.e.}, not from the anomalous modes). Apart 
from subsequent collisions with other modes producing small complex 
quartets (emerging as small bubbles in the bottom panel of Fig.~\ref{fig6}), 
the two anomalous modes remain purely imaginary, while different 
background modes generate a cascade of instabilities through the bifurcation
of real eigenvalue pairs. 

\section{Conclusions} 
In summary, we have studied the existence, stability and dynamics of dark solitons in 
spin-orbit coupled BECs. 
We developed a perturbative approach to obtain solitons on top of either 
constant or spatially modulated background density (stripe solitons).
A linear stability analysis has shown that within the considered parameter region constant background solitons are 
always stable, while stripe solitons are stable only close to the 
linear limit. The eigenfrequency of the anomalous mode associated with the oscillatory 
motion of solitons in a parabolic trap was determined analytically. Importantly, 
in a certain parameter regime where the single particle spectrum features a double well structure, 
we found a {\it second} anomalous mode, which does not exist in single- or multi-component 
BECs. Exciting this mode, we found that constant background 
solitons feature a periodic structure of moving stripes in the vicinity of the soliton core, 
reminiscent of the Kelvin modulation of vortex lines; 
for stripe solitons, we observed 
an out-of-phase oscillation of the constituent solitons.  
Our work paves the way for relevant studies in higher-dimensional settings.

{\it Acknowledgements.} We thank P. Engels for helpful comments, 
and NSF, AvH and DFG for financial support.

\end{document}